\documentclass[]{spie}  

\usepackage{amsmath,amsfonts,amssymb}
\usepackage{graphicx}
\usepackage[colorlinks=true, allcolors=blue]{hyperref}

\title{Tempestas ex machina: A review of machine learning methods for wavefront control}

\author[a]{J. Fowler}
\author[b]{Rico Landman}
\affil[a]{Department of Astronomy \& Astrophysics, University of California, Santa Cruz, CA, USA}
\affil[b]{Leiden Observatory, Leiden University, Leiden, The Netherlands}

\authorinfo{Further author information: (Send correspondence to J.F.)\\J.F.: E-mail: jumfowle@ucsc.edu}

\begin{document} 
\maketitle

\begin{abstract}
As we look to the next generation of adaptive optics systems, now is the time to develop and explore the technologies that will allow us to image rocky Earth-like planets; wavefront control algorithms are not only a crucial component of these systems, but can benefit our adaptive optics systems without requiring increased detector speed and sensitivity or more effective and efficient deformable mirrors. To date, most observatories run the workhorse of their wavefront control as a classic integral controller, which estimates a correction from wavefront sensor residuals, and attempts to apply that correction as fast as possible in closed-loop. An integrator of this nature fails to address temporal lag errors that evolve over scales faster than the correction time, as well as vibrations or dynamic errors within the system that are not encapsulated in the wavefront sensor residuals; these errors impact high contrast imaging systems with complex coronagraphs. With the rise in popularity of machine learning, many are investigating applying modern machine learning methods to wavefront control. Furthermore, many linear implementations of machine learning methods (under varying aliases) have been in development for wavefront control for the last 30-odd years. With this work we define machine learning in its simplest terms, explore the most common machine learning methods applied in the context of this problem, and present a review of the literature concerning novel machine learning approaches to wavefront control. 

\end{abstract}


\section{INTRODUCTION}
\label{sec:intro}  

In the search for and characterization of extrasolar planets, direct imaging is a powerful tool to unlock the detection of planets not available via other detection methods (i.e., planets that will not geometrically transit) and characterize them without having to disentangle the light of the planet from its host star. From the ground, direct imaging is only possible with extreme adaptive optics (exAO) where a very high level of correction over a small field of view enables high contrast imaging. Only stellar point spread functions (PSF) with extremely high Strehl ratios will enable coronagraphy to effectively null host starlight, and result in high enough contrasts to resolve the light of substellar companions. 
The scope of this review is limited specifically to wavefront control methods used within the field of ground-based adaptive optics for astronomical instrumentation, with a keen interest in this extreme adaptive optics context. (For a review of machine learning methods for wavefront sensing, see Wong, 2021 \cite{Wong2022}.)

\begin{figure} [ht]
   \begin{center}
   \begin{tabular}{c} 
   \includegraphics[width=0.9\textwidth]{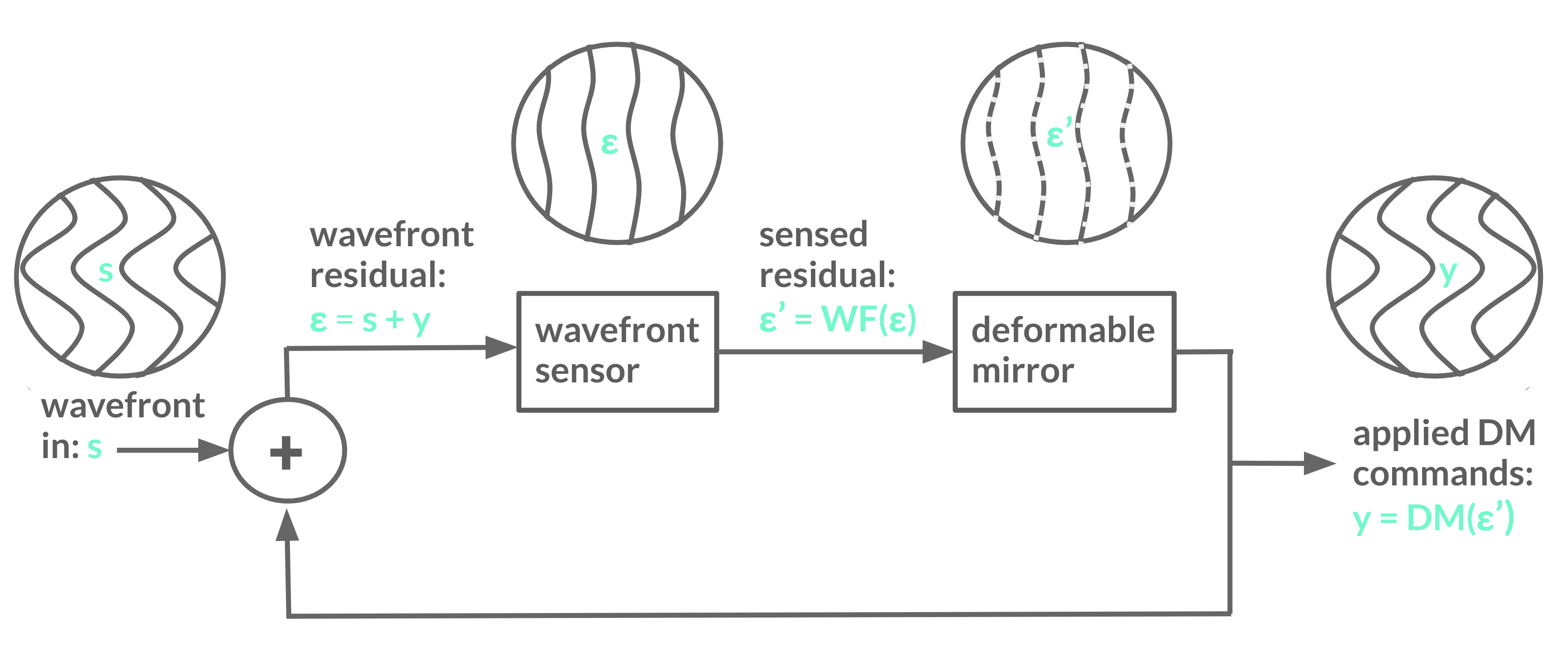}
   \end{tabular}
   \end{center}
   \caption[example] 
   { \label{fig:wfc_schematic} 
The full state of turbulence \textbf{s} enters the system and immediately encounters the wavefront correction \textbf{y}, noted by the sum junction. The wavefront sensor senses the residual wavefront $\boldsymbol{\epsilon} = \mathbf{s} + \mathbf{y}$, and reports a digitized and imperfect realization $\boldsymbol{\epsilon}'$. Finally, a computer uses this information to calculate the command update to the deformable mirror, which feeds the next iteration.}
   \end{figure}

An adaptive optics system at its simplest is made of 3 fundamental components: a wavefront sensor (which records information on the state of the wavefront after it has been aberrated by Earth-atmosphere), a deformable mirror (which attempts to correct for those aberrations), and a control algorithm that determines what correction the deformable mirror (DM) should apply given the wavefront sensor information available. Figure \ref{fig:wfc_schematic} shows a control diagram of this system.

To date, every extreme AO system runs their standard control for mid and high order modes as an integrator. This refers specifically to a closed-loop integral style controller, where the wavefront sensor is downstream of the DM, senses residual wavefront after the correction, estimates the error with a simple minimization between that residual and a flat reference, and applies that command as an addition to the current shape with some gain that is typically between $0.2<g<0.6$, or a gain optimized per mode\cite{gendron1995, males2018}. This correction is applied as quickly as possible, with the hope that the speed of the system will keep up with the speed of the evolving atmosphere. In practice this lag-time ranges from 1-5 ms.  



However, not only do we want to optimize our current observations with large telescopes like VLT, Subaru, Magellan, and Keck, as we look forward to ELTs, we will require higher fidelity correction to detect and characterize earth-like planets around sun-like stars. For example, detecting oxygen in the atmosphere of Proxima Centauri b will require a raw contrast of 3$\times 10^{-5}$ in the visible at 10 $\lambda$/D\cite{Fowler2023}, something our current control algorithms and systems have yet to accomplish. Much of this is limited by technology, but some of it is limited by the performance of the control algorithm. Furthermore, as we scale up to larger systems, with higher actuator count deformable mirrors and wavefront sensors with many more subapertures, the act of calibrating the DM to the wavefront sensor poke by poke or pixel by pixel becomes considerably more time-consuming, and a problem that is interesting to consider offloading to a machine learning algorithm. 


We expect the predictable elements of atmospheric physics are governed entirely by linear systems; state of the art atmospheric modelling for adaptive optics is done with Kolmogorov turbulence \cite{kolmogorov1941, tatarskii1961} that evolves according to the frozen flow hypothesis \cite{taylor1938}, where velocities imparted by wind layers are expected to be the driving factor. This model has been confirmed from the perspective of an AO system.\cite{poyneer2009} Additional atmospheric turbulence (i.e., boiling \cite{dravins1998}, dome seeing\cite{woolf1979}) adds random features that could still be statically encapsulated with linear noise propagation (i.e., Kalman filtering and AR models.) 


Wavefront sensing is an interesting potential source of non-linearities, covered at length in Wong, 2022\cite{Wong2022}. In short, wavefront sensing often operates as a tradeoff between sensitivity and linear sensible range (the more sensitive the wavefront sensor is, the lower the usable linear range becomes.) While we attempt to design our wavefront sensor in such a way that this assumption of linearity is valid (i.e. by introducing dynamic modulation), the Pyramid and Zernike wavefront sensors are known to have nonlinear responses for large aberrations. However, running wavefront control in closed-loop leaves residual turbulence sensed by the wavefront sensor within the linear range of most wavefront sensors used for first-stage AO. Similarly, deformable mirrors (DMs) operate with actuator pokes that have a limited linear range; in practice, the linear range of the DM is much larger than that of the wavefront sensor. 

In order to comment on and evaluation the performance of control methods, we first set priorities for current and future instruments. We define mandatory criteria and ideal exploration space for new methods within an extreme exAO system in Table \ref{tab:criteria}. 
As we summarize these methods, we examine which methods achieve some of these ideal features and whether or not they sacrifice mandatory AO requirements. 

\begin{table}[ht]
\caption{Here we define criteria deemed mandatory for extreme AO systems (i.e., any method in development cannot sacrifice these), and criteria that we could potentially gain with future technology developments. } 
\label{tab:criteria}
\begin{center}       
\begin{tabular}{|l|l|} 
\hline
\rule[-1ex]{0pt}{3.5ex}  mandatory criteria &   \\
\hline
\rule[-1ex]{0pt}{3.5ex}   & achieves at least 60\% Strehl ratio in the IR \\
\rule[-1ex]{0pt}{3.5ex}  & runs in real time  \\
\rule[-1ex]{0pt}{3.5ex}   & robust over a full night  \\
\rule[-1ex]{0pt}{3.5ex}   & accounts for static NCPAs \\
\hline
\rule[-1ex]{0pt}{3.5ex}  ideal criteria &   \\
\hline
\rule[-1ex]{0pt}{3.5ex}  & accounts for non-linearities  \\
\rule[-1ex]{0pt}{3.5ex}  & does not require high fidelity instrument model \\
\rule[-1ex]{0pt}{3.5ex}  & high sky coverage at high performance \\
\rule[-1ex]{0pt}{3.5ex}  & tunable and flexible \\
\rule[-1ex]{0pt}{3.5ex}  & accounts for \textit{dynamic} NCPAs \\
\rule[-1ex]{0pt}{3.5ex}  & does not require consistent operator attention \\
\hline
\end{tabular}
\end{center}
\end{table} 

\section{MACHINE LEARNING METHODS COMMON TO THE FIELD}
\label{sec:sections}

\subsection{Formal Definitions of Machine Learning}

Despite recent explosions in complex machine learning and artificial intelligence algorithms meant to write software, have a conversation, or win chess games, for this review a more basic definition of machine learning will suffice. We define machine learning as any problem that uses one set of data to make rules for a second set of data \cite{MLbook}. From this definition, machine learning not only includes complicated many-layer neural networks and artificial intelligence engines, but also classic linear regression problems, extrapolation problems, and many basic control engineering and statistical concepts (i.e., Kalman filtering and linear quadratic Gaussian controllers.) In this section, we will overview the most common machine learning methods applied to wavefront control, as well as comment on the overlap in methods. 
 
\subsection{Kalman Filtering and Linear Quadratic Gaussian Control}

\subsubsection{Kalman filtering}

\begin{figure} [ht]
   \begin{center}
   \begin{tabular}{c} 
   \includegraphics[width=0.5\textwidth]{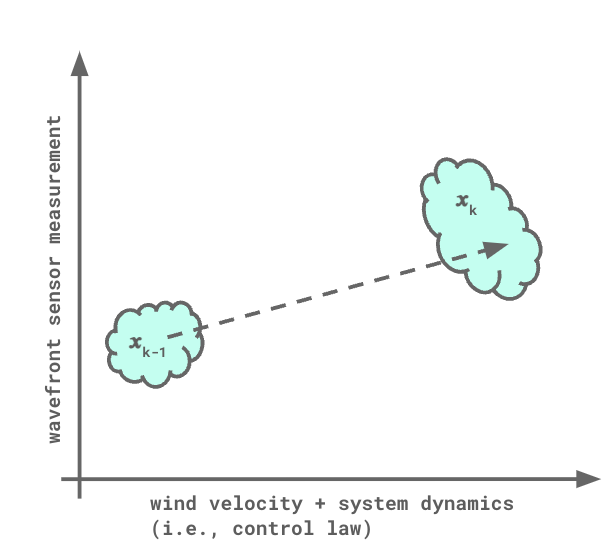}
   \end{tabular}
   \end{center}
   \caption[example] 
   { \label{fig:kalman_filter} 
“A Kalman filter is used to generate estimates of the states, given the noise-corrupted measurements.”\cite{Paschall1993} A Kalman filter projects a state vector and a Gaussian uncertainty cloud forward according to a physical law. }
   \end{figure}

Kalman filtering \cite{kalman1960} is a statistical prediction tool, common to engineering and mathematics alike. Simply stated, a Kalman filter is a robust way to predict the evolution of a property according to a physical law, a Gaussian uncertainty distribution, and data fed back into the system to verify and improve that estimate. Figure \ref{fig:kalman_filter} shows a visual representation of how some property (for our purposes, the measurement from a wavefront sensor) evolves according to some physical law (for our purposes, the wind velocity affecting the atmospheric turbulence, any dynamics in the system, and the control law we are applying), and a cloud of uncertainty that accompanies and is altered by that prediction. 

A Kalman filter starts with a state vector $\mathbf{\hat{x}}$, which encapsulates all the variables that we predict at each iteration, and a covariance matrix $\mathbf{P}$, which uses the covariance to estimate the uncertainty of those measurements. Given a physical law $\mathbf{F}$, a control law $\mathbf{B}$, and a noise term $\mathbf{Q}$, we predict the next state vector and covariance matrix with

\begin{eqnarray}
    \mathbf{\hat{x}_k} &=& \mathbf{F_k}\mathbf{\hat{x}_{k-1}} + \mathbf{B_{k}}\vec{\mathbf{u}_k} \\
    \mathbf{P_k} &=& \mathbf{F_k} \mathbf{P_{k-1}}\mathbf{F_k^T} + \mathbf{Q_k} 
\end{eqnarray}

Despite a simple step-by-step update formula, a Kalman filter is not typically updated at each iteration in practice. Instead, a steady-state solution is found after training from previous system data. 

\subsubsection{Linear quadratic Gaussian control}
\begin{figure} [ht]
   \begin{center}
   \begin{tabular}{c} 
   \includegraphics[width=0.85\textwidth]{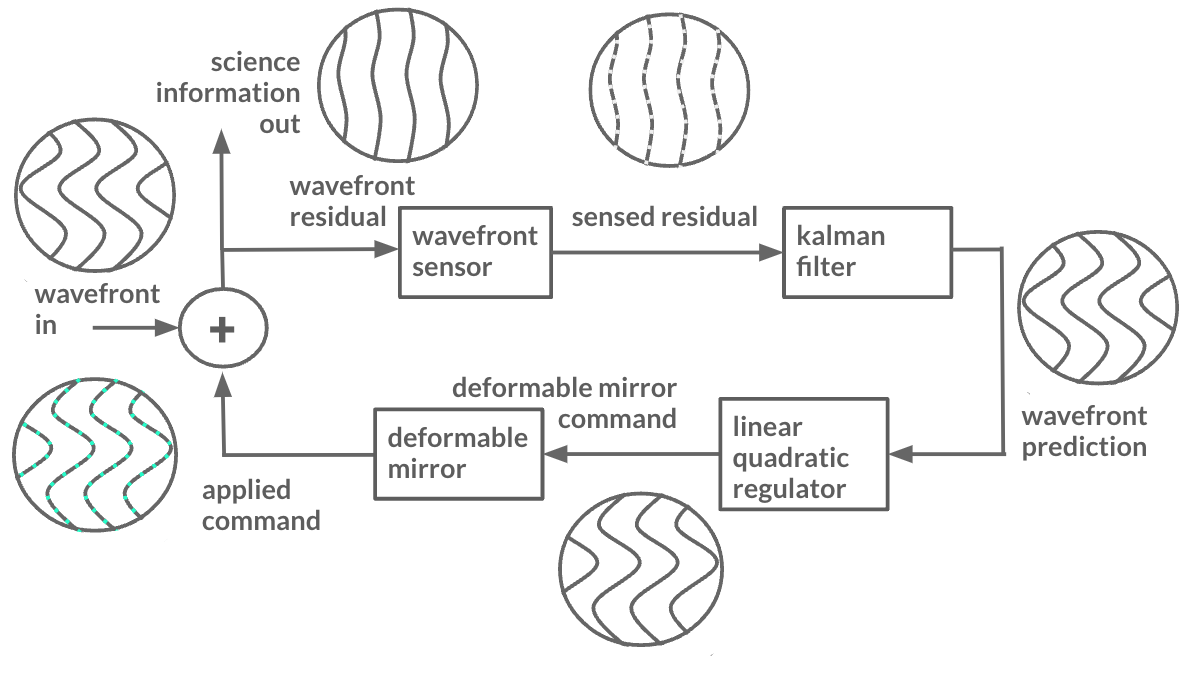}
   \end{tabular}
   \end{center}
   \caption[example] 
   { \label{fig:lqg} 
Control diagram of a linear quadratic Gaussian (LQG) controller. In this system, a sensed wavefront residual is fed to a Kalman filter, which predicts the future state of the wavefront, calculates the control law with a linear quadratic regulator, and sends that control command to the deformable mirror to correct the system. }
   \end{figure}

A common pair in control engineering is a Kalman filter and a linear quadratic regulator, or a controller that aims to minimize a quadratic cost function at each iteration. This combination is referred to as a linear quadratic Gaussian (LQG) controller; a controller that assumes Gaussian errors, predicts forward with a linear physical law, and regulates the system with a quadratic cost function. Figure \ref{fig:lqg} shows an example of an LQG controller in the context of wavefront control. 

\begin{equation}
\mathbf{J} = \mathbf{E}\left<\mathbf{\Sigma}^{N}_{i=0}\frac{1}{2}\left[\mathbf{x^T}(t_i)\mathcal{X}\mathbf{x}(t_i) + \mathbf{u^T}(t_i)\mathcal{U}\mathbf{u}(t_i)\right]\right>
\end{equation}

$\mathbf{J}$ is an example cost function \cite{Paschall1993} used for a linear quadratic regulator, given some state vector $\mathbf{x}$, deformable mirror commands $\mathbf{u}$, and expectation value $\mathbf{E}$. $\mathbf{J}$ minimizes the error in the state as well as DM commands to estimate the next DM command.

\subsection{Empirical Orthogonal Functions}

Empirical orthogonal functions (EOF)  \cite{Guyon2017} estimates the future of a wavefront by using past data to find linear temporal trends. It is basis independent (it can be used with zonal modes, Fourier models, KL modes, or any mix that allows for robust non-competing control), though in practice it has been mostly implemented with zonal (one data point per wavefront sensor subaperture/DM actuator) modes. 

First, we build a history vector \textbf{h}(t), that contains some number of associated wavefront sensor states. Depending on the implementation, this vector can be per mode (i.e. for m modes in a basis, $w_0, w_1, ... w_{m-1}$), or \textbf{h}$_m$, mapping to a specific mode. Equation 4 shows the example for all modes and many states in time. 

\begin{equation}
\mathbf{h}(t) = 
\begin{bmatrix}
\mathbf{w}_0(t) \\
\mathbf{w}_1(t) \\
... \\
\mathbf{w}_{m-1}(t) \\
\mathbf{w}_0(t -dt) \\
... \\
\mathbf{w}_{m-1}(t-(n-1)dt) 
\end{bmatrix}
\end{equation}

The goal of EOF is to build a predictive filter \textbf{F} (or \textbf{F}$_m$ if it maps to a single mode) that will predict a future wavefront, such that some \textbf{predicted wavefront} $= \mathbf{F}\mathbf{h}$. This predictive filter can be calculated every iteration, or updated every time the linear evolution of the turbulence changes (which we would roughly expect to follow evolving wind velocity over the course of a night.) 

\begin{equation}
    \mathbf{F} \mathbf{h}(t) = 
    \begin{bmatrix}
    \mathbf{w}_0(t+dt) \\
    ... \\
    \mathbf{w}_{m-1}(t+dt)
    \end{bmatrix}
\end{equation}

To find the predictive filter, we build \textbf{D}, a matrix which holds many history vectors that act as training data. They will be minimized against a truth condition \textbf{P}, which holds the correct prediction corresponding to each history vector. This leads to a simple minimization problem

\begin{equation}
\textrm{min}_\mathbf{F}||\mathbf{D}^T\mathbf{F}^T - \mathbf{P}^T||^2 .
\end{equation}

The solution for the predictive filter can be found using a simple pseudo-inverse: 

\begin{equation}
\mathbf{F} = ((\mathbf{D}^T)^\dagger \mathbf{P}^T)^T.
\end{equation}

This inversion can be done in a handful of ways; Guyon 2017 \cite{Guyon2017} used a singular value decomposition (SVD), but other works \cite{jensen2019, vanKooten2022, Fowler2022} have used a least squares inversion. 

EOF as a predictor works so simply because we consider some wavefront \textbf{w}$_m$(t) as the full state of turbulence that the atmosphere imparts. In practice, this is not the measurement we have in a typical AO system, which runs in closed-loop, where our wavefront sensor will measure a residual after some DM correction has been applied. In this implementation, EOF uses pseudo-open loop (POL) reconstructed values of the full wavefront, where information on the applied DM commands and measured wavefront sensors residuals are used to reconstruct what the full state of the turbulence would have been. This is how this is run on-sky both at Subaru/SCExAO \cite{Guyon2018} and Keck/NIRC2 \cite{vanKooten2022}. 

However, the math represented by EOF is arguably comparable to that of Dessenne, 1998\cite{Dessenne1998}, which outlined a similar method, but builds a modified history vector using DM commands and wavefront sensor slopes, instead of reconstructing the full state of the turbulence for each input mode. This implementation is more computationally expensive, but runs in closed-loop. 

\subsection{Neural Networks}
    The previous methods all rely on a linear state space model; the predicted future state can be expressed as a linear combination of previous states and commands. The advantage of this linear model is that one can often find analytical solutions that minimize a quadratic cost function, as discussed in the previous section. However, this assumption of linearity is not valid in all cases. An alternative is to use a nonlinear state-space model. This can be done using Neural Networks (NN), which can act as arbitrary function approximators. These NN's utilize multiple consecutive matrix multiplications with a so-called activation function in between. This activation function is nonlinear, allowing the NN to model nonlinear functions. These NN's are often visualized by having multiple "hidden" layers between the input and output, where the activation $\textbf{y}$ in layer $i$ can be mathematically expressed as:
    \begin{equation}
        \textbf{y}_i = f(W\textbf{y}_{i-1} + \textbf{b}).
    \end{equation}
    
    Here, $W$ are the weights of that layer, $\textbf{b}$ the bias, and $f()$ the activation function.  A visualization of a NN is shown in Fig. \ref{fig:rl_nn}. Note that a single-layer NN without an activation function reduces to a familiar linear model, such as in EOF. The amount of hidden layers and number of neurons in each layer is tweaked based on the complexity of the problem. The weights and biases can be optimized by minimizing a chosen loss function on a training dataset. This is commonly done using stochastic gradient descent based optimizers in combination with the backpropagation algorithm. Libraries such as \texttt{keras} and \texttt{pytorch} hide most of these complexities to end users, which makes building and training basic NN's much easier.

    \subsubsection{Advanced architectures}
    For complex nonlinear problems basic NN's quickly run into the curse of dimensionality, where the number of free parameters explodes. Luckily, the parameterization of these NN's can be adapted to specific problems by using domain knowledge. This can be seen as a strong form of regularization on the complexity of the model. For example, Convolutional Neural Networks (CNN's) are popular for image processing tasks. They use the knowledge that features are local and translational invariant to drastically reduce the number of free parameters that have to be optimized. For our purposes, CNN's can be used to capture spatial correlation in the wavefront. Alternatively, Recurrent Neural Networks (RNN's) are commonly used for problems involving sequences. They are well suited to deal with the temporal aspect of predictive control. Popular variants of RNN's are Long-short term memory (LSTM) networks and Gated Recurrent Units (GRU). 

   \begin{figure} [ht]
   \begin{center}
   \begin{tabular}{c} 
   \includegraphics[width=0.5\textwidth]{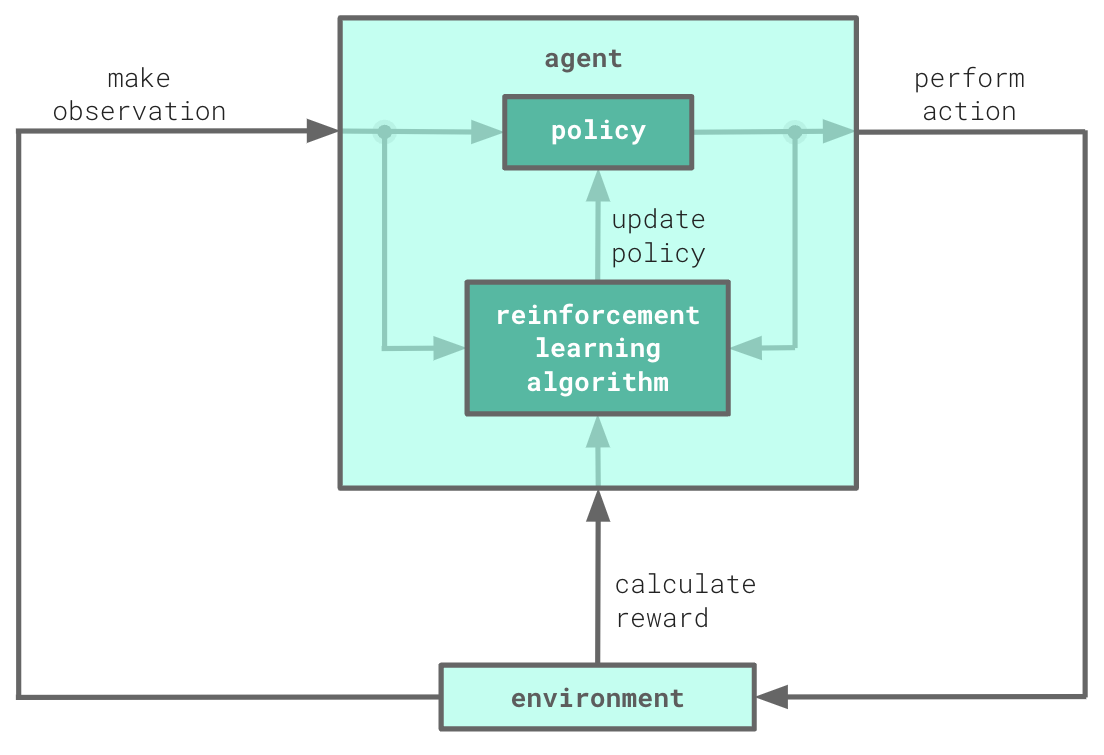}
     \includegraphics[width=0.5\textwidth]{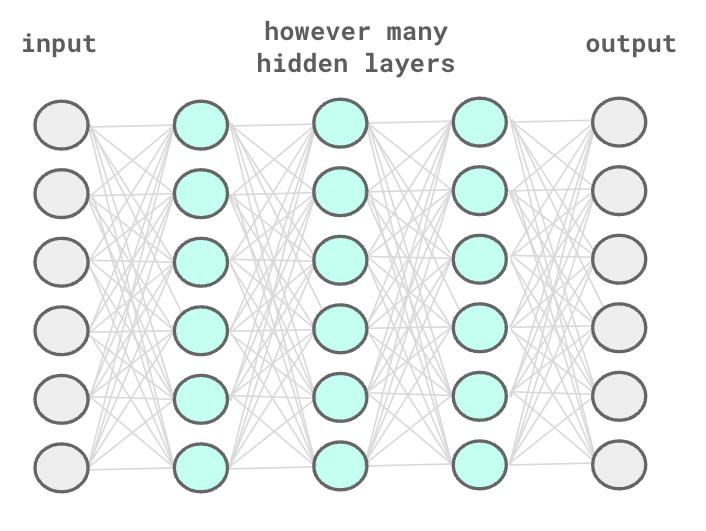}
   \end{tabular}
   \end{center}
   \caption[example] 

   { \label{fig:rl_nn} 
Left: Diagram of a reinforcement learning (RL) algorithm. An RL algorithm enacts some policy (control law) on some environment, makes observations, optimizes itself according to a reward. Right: Diagram of a neural network (NN). A neural network finds non-linear connections between layers of nodes. }
   \end{figure}

\subsection{Reinforcement Learning}
    Reinforcement Learning (RL) is a fast growing field that deals with learning optimal control through interaction with environment. The goal of these algorithms is to find the optimal commands that maximize some arbitrary optimization objective. RL algorithms can be split up into two main branches: model-free and model-based. Note that models are used in both cases.
    
    \subsubsection{Model-based reinforcement learning}
    Model-based RL algorithms explicitly learn a state-space or dynamics model. Using this state-space model and the currently measured state, they aim to find the control command which maximizes the optimization objective at each iteration. This can be done using planning algorithms, which can be quite computionally expensive to run at kHz frequencies. In the case of a linear dynamics model with a quadratic cost function, model-based RL reduces to the LQG problem with online identification of the state-space model.
    
    \subsubsection{Model-free Reinforcement Learning}
    Instead of learning a dynamics model, model-free RL algorithms aim to directly learn an optimal parameterized control law. Popular choices of model-free RL are variants of the policy gradient algorithm, such as Soft Actor-Critic (SAC) and Deep Deterministic Policy Gradient (DDPG). All these algorithms try to learn a differentiable model of the optimization metric as a function of the state and control command. One can then obtain the gradient of the cost function with respect to the control commands. This can subsequently be used to optimize the parameters of the control law. Model-free RL methods provide a parameterized control law, with which inference speeds are typically fast enough to run at kHz frequencies. A disadvantage of model-free algorithms are that they are typically less data efficient and would thus more slowly adapt to changing conditions. Promising hybrid methods, that combine the advantages of model-free and model-based algorithms, are also being developed.
    
\subsection{Overlaps Between Fields} 

 \begin{figure} [ht]
   \begin{center}
   \begin{tabular}{c} 
   \includegraphics[width=0.7\textwidth]{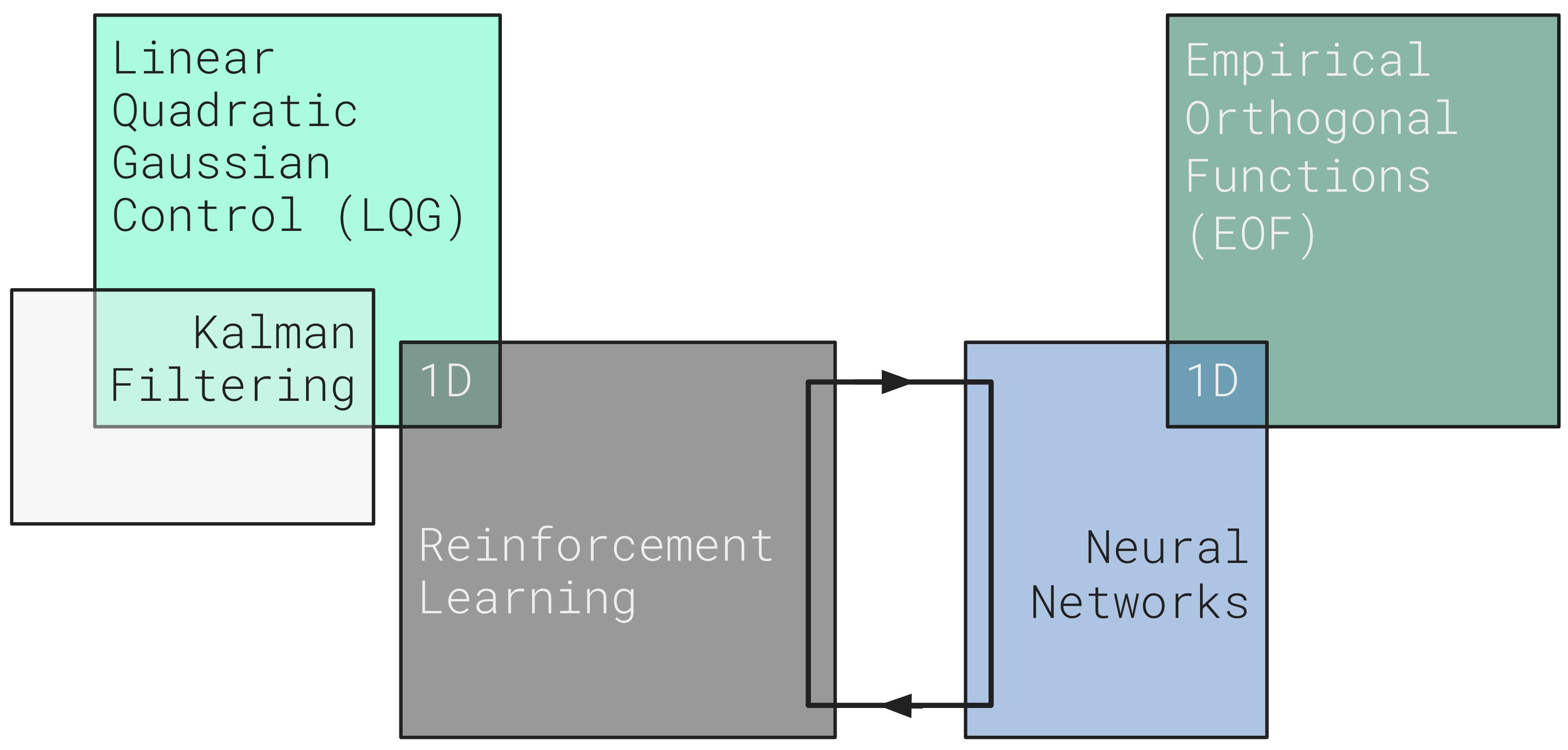}
   \end{tabular}
   \end{center}
   \caption[example] 

   { \label{fig:rl_nn} 
Reinforcement learning is a particular way to frame a problem; a linear example of reinforcement learning is a linear quadratic Gaussian controller (which requires a Kalman filter as its estimator.) Neural networks are often used to solve reinforcement learning problems, and empirical orthogonal functions is equivalent to a neural net that has been constrained to a single layer and temporal evolution. }
   \end{figure} 

In summary, four major concepts will be discussed in the rest of this review. Reinforcement learning and neural networks are two common elements of machine learning that often go hand-in-hand: reinforcement learning as a particular problem type, and neural networks as a common solver for that problem type.

Narrowing the scope of these problems and making them linear shows that an LQG controller is simply a reinforcement learning problem with more conditions about its error propagation, physical law, and cost function. In particular, a mathematical analysis of the overlap between a reinforcement learning problem and an LQG controller is available in Recht, 2018\cite{RLvsLQG}. Kalman filtering is often utilized as an element of an LQG controller, though one of the methods we describe is closer to a pure Kalman filter approach. Finally, EOF acts as a one-layer neural network, and works that have compared EOF to neural networks\cite{Wong2021, Hafeez2022} find that they converge to the EOF solution for linear problems.

\section{RESULTS FROM THE LITERATURE}

\subsection{Linear Methods}

\subsubsection{Empirical orthogonal functions}

Arguably the first predictive controller for astronomical telescopes is from a work by Dessenne, 1998\cite{Dessenne1998}. This uses a method quite similar to empirical orthogonal functions, except that it runs in closed-loop; instead of associating the full state of the wavefront in time it builds a similar history vector that includes both wavefront sensor residuals and DM commands from every iteration. The truth condition that the filter is optimized against uses the rejection transfer function (i.e., a high fidelity pseudo-open loop reconstruction) to reconstruct the full wavefront from the DM commands and wavefront sensor slopes. This method was implemented on-sky in 1999\cite{Dessenne1999}, at the 1.52 meter visible telescope at Haute Provence. They cite a 30$\%$ relative Strehl increase over a classic integrator, which maps to an increase of $\sim 11\%$ to $14\%$ Strehl over varying $r_0$ conditions at a central wavelength of 650 nm. 

After Guyon proposed EOF in 2017\cite{Guyon2017}, it was implemented on the 8 meter Subaru/SCExAO in 2018\cite{Guyon2018}. An example of predictive control on/off shows visual improvement when comparing two on-sky images, and provides a factor of up to 3 improvement in contrast at close-in inner working angles\cite{currie2019}.

Shortly thereafter, EOF was also applied in simulation to the 8 meter VLT/SPHERE telemetry\cite{vanKooten2020}. This application used telemetry data (in the form of wavefront sensor slopes and DM commands) to apply predictive control as if it were a post-processing method, and compare this to the actual on-sky performance of the integrator. van Kooten, 2020 found a 5.1 factor of improvement in RMS error over the residuals of the SAXO integrator, and estimated an additional 2.1 increase over an idealized AO system for VLT/SPHERE when applied to data taken at a central wavelength of 658 nm. 

Finally, work started in 2019\cite{jensen2019} to apply this on-sky at the 10 meter Keck observatory. Performance improvements have been shown with a similar post-processing telemetry application\cite{jensen2019, Fowler2022}, and van Kooten, 2022\cite{vanKooten2022} demonstrates an on-sky application. van Kooten, 2022 found a factor of 3 contrast improvement over the classic integrator at an inner working angle of 3-7 $\lambda$/D, and a 1.5 factor decrease in standard deviation when applied to L band coronagraphic observations, making the predictive controller consistently more robust than the integrator. A similar test also looked at Strehl from the Bracket $\gamma$ narrowband filter, and found a Strehl increase from $\sim19\%$ with the integrator to $\sim23\%$ with the predictive EOF method. 


\subsubsection{Kalamn filtering}

Haffert, 2021\cite{Haffert2021} takes a Kalman filtering-like approach, with a subspace predictive controller. Subspace predictive control builds and updates a predictive Kalman filter, with a state vector built of wavefront sensor slopes and DM commands; this work could also be considered an extension of the Dessenne, 1998\cite{Dessenne1998} work, but without the need for rejection transfer function (or pseudo-open loop reconstructed) data to train against. This method, when applied to the MagAO-X daytime bench (an instrument for the 8 meter Magellan-Clay telescope) shows a contrast improvement by a factor of of up to 10-20 for low wind speeds and up to two orders of magnitude for higher wind speeds at a central wavelength of 760 nm \cite{Haffert2021}.


\subsubsection{Linear quadratic Gaussian controllers}

Linear quadratic Gaussian control was suggested as a controller for astronomical telescopes in 1993\cite{Paschall1993}, making 30 years of machine learning methods for wavefront control with the publication of this proceeding. The classic implementation of this method quickly gained popularity for the control of lower order modes. LQG runs as the standard low order controller for the VLT/SPHERE, where low order LQG performed with an RMS error of 0.87 mas on tip and tilt, as compared to a classic integrator at 12.8 mas\cite{Petit2011}. The Gemini Planet Imager (GPI) also uses low order LQG control, in particular to control consistent and specific frequencies; for example one 27 Hz frequency induced a 19.5 mas tilt, and with the LQG 2.3 mas of tilt remained at a central wavelength of 1570 nm\cite{Poyneer2016}.

The only application of this method as a full controller on-sky was on the 4.2 meter CANARY AO demonstrator at the William Herschell Telescope, which reports a Strehl of $50.9\%$ for a classic integrator, and $61.9\%$ for the LQG controller in the H band\cite{Sivo2014}.
A reformulated data-driven LQG\cite{sinquin2020} was also tested on CANARY/WHT, finding up to $7\%$ Strehl improvement (from 0.067 to 0.051 radians RMS) when run on-sky in the H band.  

Predictive Fourier control (PFC)\cite{Poyneer2007} is another LQG alternative, which learns wind speeds from telemetry data (often called a wind-driven predictor). Its original implementation was for GPI-like simulations, which showed an integrator residual RMS error of 37.2 nm, as compared  to 21 nm for the predictive Fourier control\cite{Poyneer2007}. Fowler, 2022 applied this method to Keck telemetry, finding an improvement in RMS error on by a factor of 2.5 when comparing PFC to on-sky integrator residuals\cite{Fowler2022}. Most recently lab results from the LLAMAS (Low Latency Adaptive Optical Mirror System) testbed show a decrease in temporal error power by up to 3, with varying wind speeds and spatial frequencies\cite{Poyneer2023}. The predictive Fourier control method also allows for non-integer lags in the predictor, which makes for a highly tunable predictive corrector.

\subsection{Neural Networks}
    \subsubsection{Wavefront prediction}
    Some works have explored the use of neural networks (NN) for predicting the evolution of atmospheric turbulence or wavefront sensor telemetry \cite{Liu2020,Wong2021, Hafeez2022}. These mainly focused on the prediction of (pseudo-)open loop measurements, and all show promising improvements over classical, non-predictive controllers (i.e., an integrator). We note that these methods build a nonlinear state-space model that predicts the evolution of turbulence, but do not build an optimal nonlinear controller; a classic integrator is often used as controller. From these works, Wong, 2021\cite{Wong2021} used a simple feed-forward NN, while Hafeez and Liu used Long-Short Term Memory networks to deal with the temporal aspect of the data\cite{Hafeez2022, Liu2020}. 
     
    Swanson, 2022 \cite{Swanson2021} used NN prediction in a closed-loop AO system, showing that in closed-loop there can be a mismatch between the training distribution and observed distribution, which can be mitigated by using an adverserial prior. Swanson, in particular, explored the role of guidestar magnitude and increased sky coverage, showing that the NN method was able to show major performance improvements for a 16th magnitude guidestar, from a Strehl ratio of $40\%$ with an integrator to $70\%$ with the NN, for a simulated 8 meter telescope in the R band. 

    \subsubsection{Reinforcement learning}
    
    The challenge with nonlinear state-space models is that it is not trivial to find an analytical solution to the control problem with a quadratic cost function (unlike linear examples). To that end, reinforcement learning (RL) aims to derive optimal nonlinear controllers by e.g. using planning algorithms (model-based) or through direct optimization of a parameterized control law (model-free). An additional feature of RL algorithms is that one is not limited to a quadratic cost function of the state variables but is free to choose the optimization objective (i.e., reward), allowing one to directly optimize for e.g. Strehl or contrast.

    Nousiainen, 2021\cite{Nousiainen2021} used model-based RL in a simulated 24x24 AO system with a Shack-Hartmann and demonstrated an improvement in Strehl and contrast compared to an integrator, showing that their algorithm is also able to deal with DM to wavefront sensor misregistrations, but notes that the algorithm may be too computationally expensive to run at kHz frequencies. A consecutive work (Nousiainen, 2022\cite{Nousiainen2022}) solved this issue by using a parameterized control law; they not only show significant contrast improvements in a simulated VLT-like system with a PWFS, and decreased speckle variance in the lab with MagAO-X, but also run a simulations of a 40 meter telescope in real time with this controller, increasing performance from $88\%$ Strehl with a classic integrator to $90\%$ Strehl for the RL method, calculated at 1650 nm. Most recently, Nousiainen demonstrated these results the GHOST testbench and showed promising initial results in a cascaded AO system.\cite{Nousiainen2022b}

    On the other hand, in the work of Landman, 2020, and 2021\cite{Landman2020, Landman2021}, a model-free algorithm is chosen based on the Deep Deterministic Policy Gradient algorithm. Landman shows improvements in vibration mitigation in simulations and lab experiments and a large improvement in contrast in a simulated high-contrast imaging instrument on an 8 meter telescope (from $10^{-4}$ for the integrator to $2\times10^{-5}$ for the RL method, calculated at 1000 nm). By utilizing a combination of an LSTM and CNN, the RL algorithm built a policy that adjusts to changing atmospheric conditions without online learning, making the controller highly adaptable to changing on-sky conditions. 
    
    Pou, 2022 \cite{pou2022} also uses a model-free approach using the Soft Actor-Critic algorithm and considers a distributed approach using multi-agent RL. Pou shows improvements in Strehl for a simulated 8-meter telescope with Shack-Hartmann WFS and PWFS, from $85\%$ Strehl to $91\%$ Strehl calculated at 1650 nm.\cite{Pou2022b} However, for the Pyramid, this performance required a nonlinear wavefront reconstructor before the RL controller. 

\subsection{Are Non-Linear Methods Justified}

Despite pointing to non-linearities as a potential goal for new wavefront control algorithms, some interesting work in the field has explored whether or not temporally evolving non-linear controllers are necessary. Multiple works \cite{Hafeez2022}, \cite{Wong2021} have compared classic linear solvers (i.e., empirical orthogonal functions or a neural net with a single layer) to multi-layer approaches, only to find that when considering temporally evolving statistics the multi-layer systems converge to the same answers, sometimes with added noise in the multi-layer networks. 

Haffert, 2021 \cite{Haffert2021} argues that running in closed-loop maintains the condition of linearity for both Pyramid wavefront sensors (PWFS) and DM actuators, meaning non-linear reconstructions and system dynamics are not worth encapsulating in temporally evolving models. However, other works \cite{Nousiainen2022, Landman2021} demonstrate that a non-linear solver can learn the entire system model, including wavefront sensing response, deformable mirror dynamics, and DM to wavefront sensor misregistration, which leads to a more painless and flexible system implementation. Furthermore, Wong 2021 \cite{Wong2021} found that despite converging to the same solution, the neural network found the answer with $\sim$5 seconds of training data, as compared to a linear predictor which needed $\sim$40 seconds of training data.


\section{CONCLUSION}

\begin{figure} [h!]
   \begin{center}
   \begin{tabular}{c} 
   \includegraphics[width=0.8\textwidth]{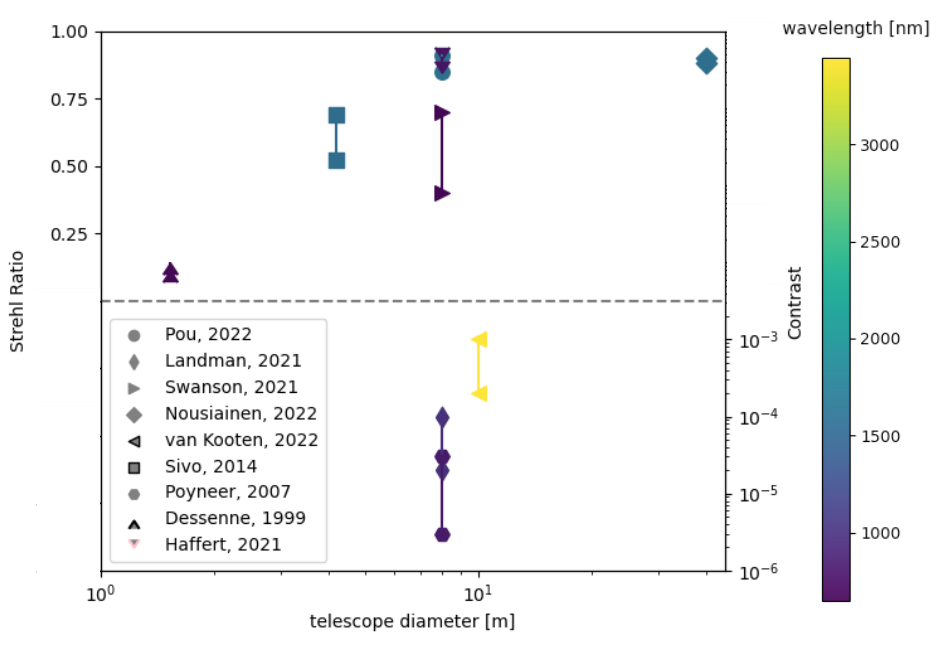}
   \end{tabular}
   \end{center}
   \caption[example] 

   { \label{fig:comparative_performance} Comparative performance of machine learning methods from the literature with increasing telescope diameter, connecting integrator performance to machine learning method performance. Everything above the dashed line is a reported Strehl ratio, and below a contrast. Color indicates wavelength. No outline: simulation, pink outline: lab results, and black outline: on-sky testing. Only van Kooten, 2022\cite{vanKooten2022} was for a large telescope with a coronagraphic imager.}
   \end{figure}

With the 1993 Paschall \& Anderson paper \cite{Paschall1993} on LQG controllers for astronomical telescopes, we celebrate 30 years of machine learning for wavefront control, over which machine learning approaches to wavefront control have been suggested, simulated, tested in the lab, and rarely tested on sky. A variety of methods show promising results, but many of these control algorithms go no further because they have yet to be consistently verified on-sky. Figure \ref{fig:comparative_performance} compares performance from the described methods, either in reported Strehl ratio, or contrast. This is not a perfect comparison, as these performance reports span different wavelength regimes, actuator counts, wavefront sensors, and telescope diameters. 

As we design the next generation of extremely large telescopes, only a few \cite{Guyon2018,vanKooten2022} works show proven results on-sky (for a 8-10 meter class telescope controlling all modes with a ML method), making it difficult to promise or benchmark performance for future systems. Furthermore, most of these efforts focus on a single algorithm and its development and progression on a single system. Few efforts \cite{Fowler2022, Wong2021, Hafeez2022} have attempted to compare the performance of these methods side-by-side, to determine how these methods may improve performance, stability, or make for simpler problems across a uniform simulation, lab demonstration, or telescope system.

\appendix    

\section{MATH MISTAKES IN FUNDAMENTAL PAPERS}
\label{sec:spite_appendix}

\subsection{Paschall \& Anderson, 1993}

The first paper to describe applying an LQG controller to an astronomical telescope system. The quadratic cost function described in Eq (17) of the original paper \cite{Paschall1993}, written as: 
\begin{equation}
\mathbf{J} = \mathbf{E}\left<\mathbf{\Sigma}^{N}_{i=0}\frac{1}{2}\left[\mathbf{x^T}(t_i)\mathcal{X}\mathbf{x}(t_i + \mathbf{u^T}(t_i)\mathcal{U}\mathbf{u}(t_i)\right]\right>
\end{equation}
is missing a close paren, it should read:
\begin{equation}
\mathbf{J} = \mathbf{E}\left<\mathbf{\Sigma}^{N}_{i=0}\frac{1}{2}\left[\mathbf{x^T}(t_i)\mathcal{X}\mathbf{x}(t_i) + \mathbf{u^T}(t_i)\mathcal{U}\mathbf{u}(t_i)\right]\right>
\end{equation}
This is an arguably obvious mistake, as the vector $\mathbf{x}$ must apply to some discrete timestep. 

\subsection{Dessenne et al., 1998}

Many credit Dessenne, 1998 \cite{Dessenne1998} as the first paper to describe a fully predictive controller. It builds a state vector $\boldsymbol{\phi}$, which holds both wavefront sensor residuals and deformable mirror commands. Equation (20) of the original work lays out the state vector and the predictive filter as: 
\begin{gather*}
\boldsymbol{\theta} = [b_1 ... b_{p-1}, a_0 ... a_{q-1}]^T \\
\boldsymbol{\phi}(i) = [y(i-1) ... y(i-p+1)] \\
(y-e'_{meas})(i-2) ... (y-e'_{meas})(i-1-1)]
\end{gather*}
This expression has an extra bracket inserted. It should read: 
\begin{gather*}
\boldsymbol{\theta} = [b_1 ... b_{p-1}, a_0 ... a_{q-1}]^T \\
\boldsymbol{\phi}(i) = [y(i-1) ... y(i-p+1), (y-e'_{meas})(i-2) ... (y-e'_{meas})(i-1-1)]
\end{gather*}
The entirety of the second two lines of the originaly Equation (20) make up a single expression for the state vector. This error has high potential for confusion as one could interpret it as a third expression, missing its definition, or two series of terms that should be multiplied as cross terms. 
 
\subsection{Poyneer et al., 2007}

Poyneer, 2007 \cite{Poyneer2007} is one of two major predictors explored today. It makes use of a particular modal basis, complex Fourier modes. The expression for complex Fourier modes in the original work from Equation (1) is written: 
\begin{equation}
    f_{k,l}[x,y] = \cos\left(i2\pi\frac{kx + ly}{N}\right) + i\sin\left(i2\pi\frac{kx + ly}{N}\right)
\end{equation}
which has two extra imaginary terms inserted. It should read: 
\begin{equation}
    f_{k,l}[x,y] = \cos\left(2\pi\frac{kx + ly}{N}\right) + i\sin\left(2\pi\frac{kx + ly}{N}\right)
\end{equation}
Those familiar with complex modes would likely spot this mistake immediately, but due to the waning use of complex modes it may be easy to forget where the imaginary terms lie. It should also be noted that the $\cos$ and $\sin$ of imaginary numbers provides a real answer, so it is not immediately obvious in debugging results mathematically that this is incorrect. 




\acknowledgments 
 
Thanks to Don Gavel, Sebastiaan Haffert, Maaike van Kooten, and Max Marion, who helped me better understand machine learning and control theory concepts. 

\bibliography{main} 
\bibliographystyle{spiebib} 

\end{document}